\pdfoutput=1
\documentclass[a4paper]{article}

\usepackage{INTERSPEECH2021}
\usepackage{times}
\usepackage{epsfig}
\usepackage{graphicx}
\usepackage{amsmath}
\usepackage{amssymb}
\usepackage{makecell}

\usepackage{subcaption}
\usepackage{float}
\usepackage{tablefootnote}
\usepackage{mathrsfs}
\usepackage{pifont}
\newcommand{\xmark}{\ding{55}}%
\newcommand{\cmark}{\ding{51}}%
\usepackage{setspace}
\usepackage{xcolor} 
\usepackage[hang,flushmargin]{footmisc}

\title{Active Speaker Detection as a Multi-Objective Optimization with Uncertainty-based Multimodal Fusion}
\name{Baptiste Pouthier$^{1,2}$, Laurent Pilati$^1$, Leela K. Gudupudi$^1$, Charles Bouveyron$^2$, Frederic Precioso$^2$}
\address{
  $^1$NXP Semiconductors, France\\
  $^2$Universit\'e C\^ote d'Azur, Inria, CNRS, LJAD, I3S, Maasai, France}
\email{\{baptiste.pouthier, laurent.pilati, leela.k.gudupudi\}@nxp.com \\ \{charles.bouveyron, frederic.precioso\}@univ-cotedazur.fr}

\begin{document}

\maketitle
\begin{abstract}
It is now well established from a variety of studies that there is a significant benefit from combining video and audio data in detecting active speakers. However, either of the modalities can potentially mislead audiovisual fusion by inducing unreliable or deceptive information. This paper outlines active speaker detection as a multi-objective learning problem to leverage best of each modalities using a novel self-attention, uncertainty-based multimodal fusion scheme. Results obtained show that the proposed multi-objective learning architecture outperforms traditional approaches in improving both mAP and AUC scores. We further demonstrate that our fusion strategy surpasses, in active speaker detection, other modality fusion methods reported in various disciplines. We finally show that the proposed method significantly improves the state-of-the-art on the AVA-ActiveSpeaker dataset.
\end{abstract}

\section{Introduction}

Active Speaker Detection (ASD) contemplates on identifying active speakers in a video by analyzing both visual and audio features. Hence, ASD is inherently multimodal in nature, where video and audio data are essential attributes. In recent years there has been considerable interest in the ASD methods based upon audio-visual cues \cite{871073, whos_speaking?_Chakravarty1, Chakravarty2, audio_visualcotraining_Chakravarty3, syncnet}. Despite this interest, the lack of large-scale in-the-wild datasets impeded scientific progress in the field. Unfortunately, most of the prior works are challenged by skewed results owing to the poor quality of the considered datasets.

\begin{figure}[t]
\centering
   \includegraphics[trim=0mm 4mm 0mm 0mm, clip, width=1\linewidth]{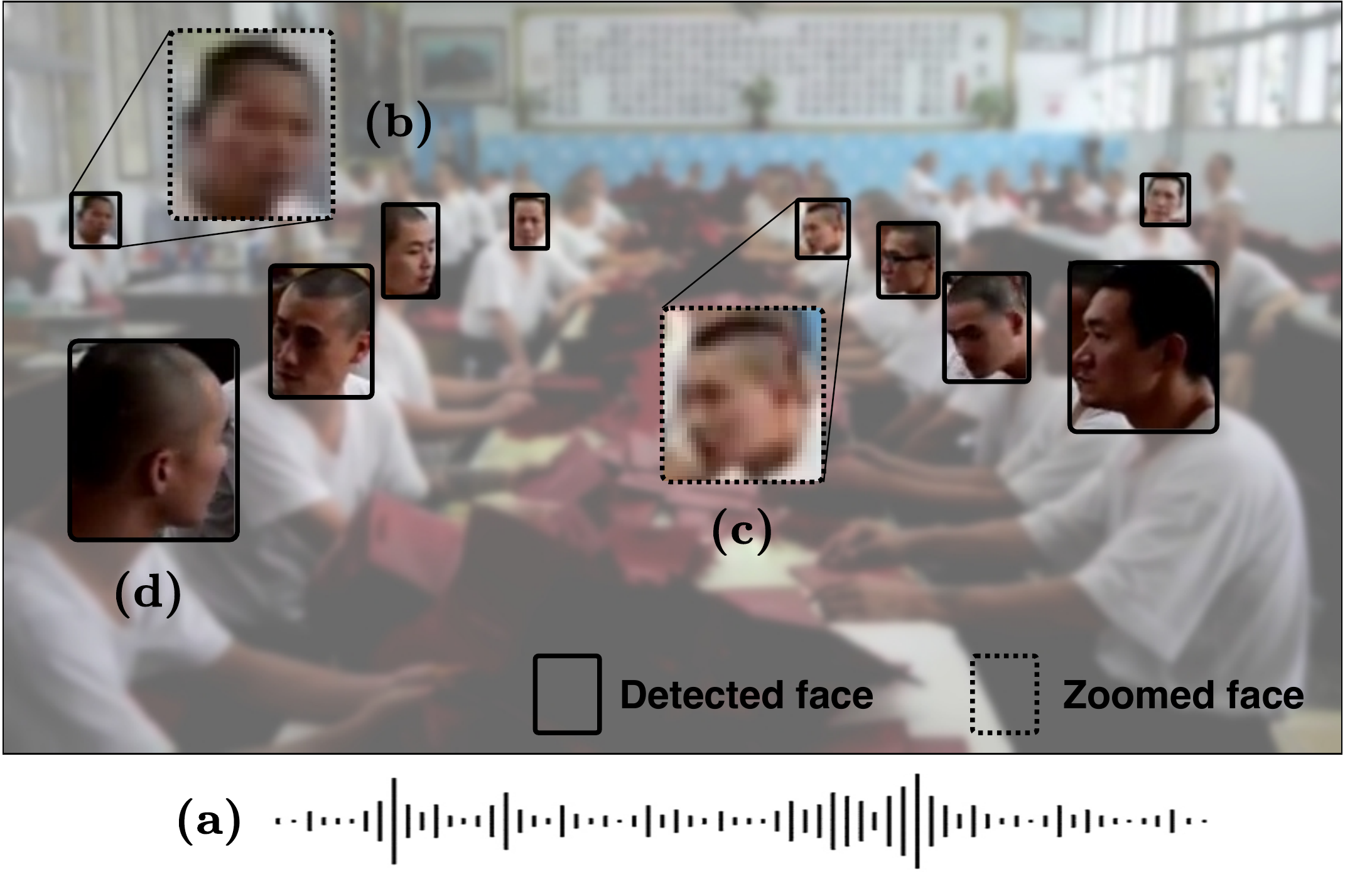}\\[-1ex]
   \caption{This is an illustration of how ambiguous both video and audio modalities are within the ASD problem. (a) represents the scene's audio track that contains speech. But the uncertainty here is who is speaking? It is difficult to say it from video also, because some face resolutions are insufficient (b,c) or characters lips may be partially (c) or entirely (d) concealed.}~\\[-4ex]
\label{fig:1}
\label{fig:mainPage}
\end{figure}

The recent AVA-ActiveSpeaker dataset \cite{AVA} can potentially overcome these limitations and reshape the ASD field of study. Together with the large in-the-wild dataset, the authors presented a baseline model based on a two-stream network that merges video and audio modalities in an end-to-end fashion. Within the annual AVA-ActiveSpeaker challenge \cite{activitynet_challenge}, Chung \cite{naver} and Zhang et al. \cite{Zhang2019MultiTask_3Dconv} improved this baseline using hybrid 3D-2D CNNs pre-trained on large-scale multimodal datasets \cite{syncnet_embedding}. Unfortunately, this approach encounters two major challenges in practice, as illustrated in Fig.\ref{fig:mainPage}: (1) multi-speaker scenario where multiple persons in a video frame are speaking, and (2)~low-resolution and/or indiscernible faces in video frames. In \cite{Alcazar_2020_CVPR}, Alc\`{a}zar et al. addressed the multi-speakers scenario by learning long-term relationships between speakers, and Huang et al. \cite{optical_flow} handled the uncertainty in video modality by adding optical flow to raw pixel representation to strengthen face characterisation. Nevertheless, these studies focused on ad-hoc objectives with limited scope and little attention has been paid to the aggregated approaches which exploit correlation between both the challenges. In this paper, we propose to learn this correlation using a cross-modal fusion that involves simultaneous learning of the uncertainty in both the modalities using a multi-objective learning scheme as described in Section~\ref{problem_formulation}.

Traditionally, in multi-task learning, uncertainty is handled by learning adaptive weighting between each task's loss functions \cite{happy, kendall2017multi}. In the field of Automatic Video Description, researchers have investigated attention-based \cite{dot_product_self_attention} fusion mechanisms that capture the importance of each of the modalities. Hore et al. \cite{fusion_before_melissa} proposed a model that selectively uses features from different modalities. This approach was later improved using hierarchical attention fusion in \cite{melissa}. Despite the growing interest of these fusion mechanisms in other disciplines, to the best of our knowledge, no studies have been conducted on ASD problem. The present paper aims to investigate different fusion schemes to solve the ASD problem. We also propose a novel audiovisual fusion mechanism as a first attempt to enrich the self-attention model with uncertainty information to disentangle the practical challenges.

The proposed method consists of learning a self-attention \cite{dot_product_self_attention} and uncertainty-based fusion mechanism that weights video and audio embeddings in an end-to-end fashion. We use the typical two-stream DNN architecture from the literature \cite{AVA, Alcazar_2020_CVPR}, with a stream per each modality, to encode both embeddings. By characterizing uncertainty of each modality, we aim at disentangling the problem presented in Fig.\ref{fig:mainPage}. Indeed, the fusion scheme we propose determines the viability of each modality at any given time to learn a comprehensive understanding of every situation towards ASD disambiguation. Our solution significantly outperforms state-of-the-art in the AVA-ActiveSpeaker dataset by 4.8$\%$ and 1.7$\%$ on validation and test sets, respectively.

\begin{figure}[t]
\centering
  \includegraphics[trim=2mm 3mm 6mm 3mm, clip, width=1\linewidth]{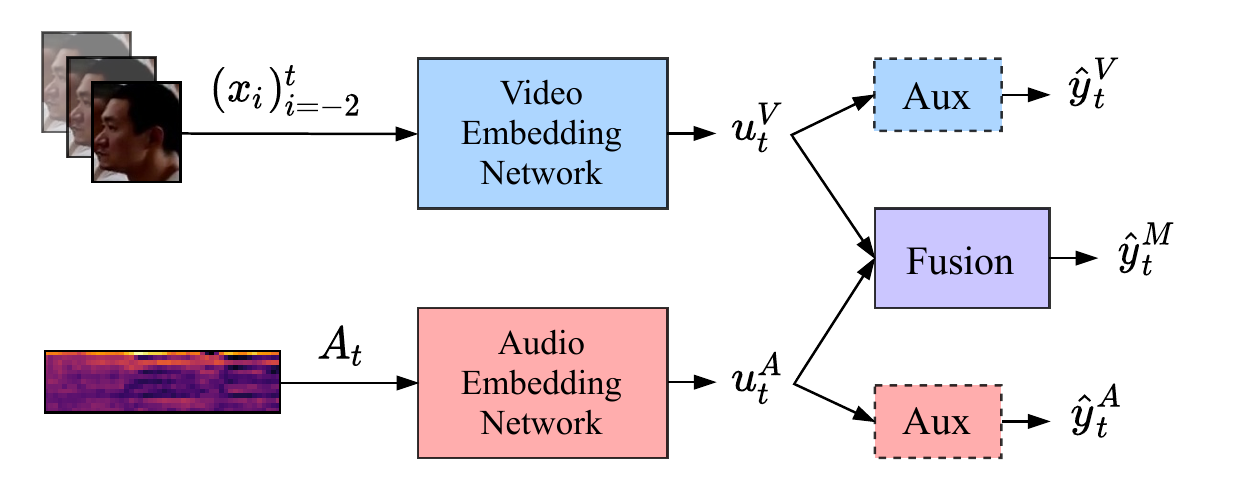}
\label{network}
\\[-3ex]
\caption{End-to-end multi-objective audiovisual network.
Video and Audio Embedding Networks are detailed in Fig.\ref{fig:embeddings}. 
"Aux" block represents an auxiliary classifier composed with two fully connected layers of 128-dim and 2-dim.}~\\[-4ex]
\label{fig:network}
\end{figure}

\section{Active Speaker Detection} \label{ASD}

Fig.\ref{fig:mainPage} demonstrates video and audio modalities may conflict, with non-talking faces affiliated with speech-labeled audio. Conflicting video and audio information lead to a compromised audio-visual learning. Most recent ASD studies \cite{AVA, naver, Zhang2019MultiTask_3Dconv, Alcazar_2020_CVPR, optical_flow} defined a trade-off optimizing both video and audio streams towards video label, sometimes using auxiliary classifiers to increase the  discriminative power of the individual streams towards the unique objective \cite{AVA, Alcazar_2020_CVPR}. However, this approach undermines the learning of the audio stream. In this section, we detail the multi-objective optimization of the video and audio streams towards their respective ground-truth labels to learn unbiased and accurate representations of video and audio modalities. Then, we introduce a novel audiovisual fusion mechanism that uses self-attention and uncertainty indicators to better disentangle the ambiguous scenarios.



\subsection{Multi-Objective Learning} \label{problem_formulation}

Let $(X_t=\{x_t^n\}_{n\in\mathbb{N}})_{t\in\mathbb{N}}$ be a series of consecutive frames, where $x_t^n$ denotes the $n^{th}$ face detected in the $t^{th}$ frame, and $(A_t)_{t\in\mathbb{N}}$ is the corresponding audio segment.
Each audiovisual sequence is noted ${M_t^n=(x_t^n,A_t)_{t\in\mathbb{N}}}$. During training, each multimodal sample $M_t^n$ is associated with a video-based ground-truth label $y_t^{nV}$ where $y_t^{nV}=1$  if the face $x_t^n$ is speaking and $y_t^{nV}= 0$ otherwise. To cater multi-objective learning, we also define audio-based ground-truth by aggregating video-based labels of each frame using Eq.\ref{eq_audio_label}.
\begin{equation}
y_t^A = \left\{
\begin{aligned}
\text{0} & \thickspace\thickspace\thickspace \text{iif} \thickspace\thickspace\thickspace \text{$\sum\nolimits_{n} y_t^{nV} = 0$},\\
\text{1} & \thickspace\thickspace\thickspace \text{otherwise}
\end{aligned}
\right.
\label{eq_audio_label}
\end{equation}
\indent We train a standard two-stream architecture, presented in Fig.\ref{fig:network}, that leverages auxiliary classifiers (Aux) to encourage video and audio networks to minimize their own loss function, thus optimizing both modalities intermediate feature representations. 
The outputs of the Video, Audio and Multimodal networks are denoted as $\hat{y}^V$,$\hat{y}^A$, and $\hat{y}^M$ respectively; we formulate these quantities as ${\hat{y}_t^{nV}=f^{W_V}(x_t^n)}$, ${\hat{y}_t^{A}=f^{W_A}(A_t)}$, and ${\hat{y}_t^{nM}=f^{W_M}(M_t^n)}$ where $W_V$, $W_A$, and $W_M$ are the weights of the respective networks. To train these weights, we define the loss function $\mathcal{L}$ as a cross-entropy loss function ${\mathcal{L}(y,\hat{y})=-\sum_{i}y_i log(\hat{y}_i)+(1-y_i) log(1-\hat{y}_i)}$. The final multi-objective loss-function $\mathcal{L}_f$ is formulated as ${\mathcal{L}_f = \mathcal{L}_M + \mathcal{L}_V + \mathcal{L}_A}$, where $\mathcal{L}_M=\mathcal{L}(y^V,\hat{y}^M)$, ${\mathcal{L}_V=\mathcal{L}(y^V,\hat{y}^V)}$, and $\mathcal{L}_A=\mathcal{L}(y^A,\hat{y}^A)$ respectively.

\begin{figure}[t]
\centering
    \includegraphics[trim=6mm 5mm 6mm 6mm,clip, width=1\linewidth]{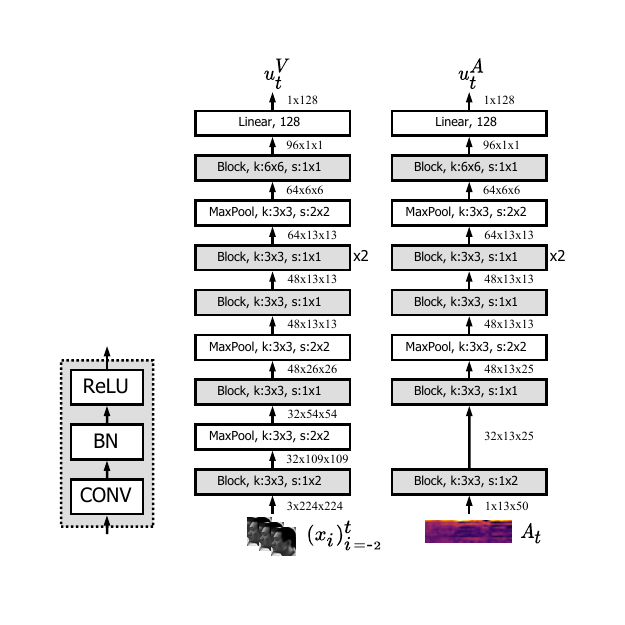}
    \\[-3.75ex]
  \subfloat[\label{embeddings:block} Block]{\hspace{.17\linewidth}}\hspace*{5mm}
  \subfloat[\label{embeddings:video} Video Network]{\hspace{.35\linewidth}}\hspace*{3mm}
  \subfloat[\label{embeddings:audio} Audio Network]{\hspace{.35\linewidth}}\hspace*{3mm}
   
\caption{(a) presents the building block of both video (b) and audio (c) embedding networks.}~\\[-4.5ex]
\label{fig:embeddings}
\end{figure}

\subsection{Cross-Modality Fusion}\label{fusion}

\begin{figure}[t]
\centering
  \includegraphics[trim=21.3mm 13.8mm 20mm 29mm,clip, width=1\linewidth]{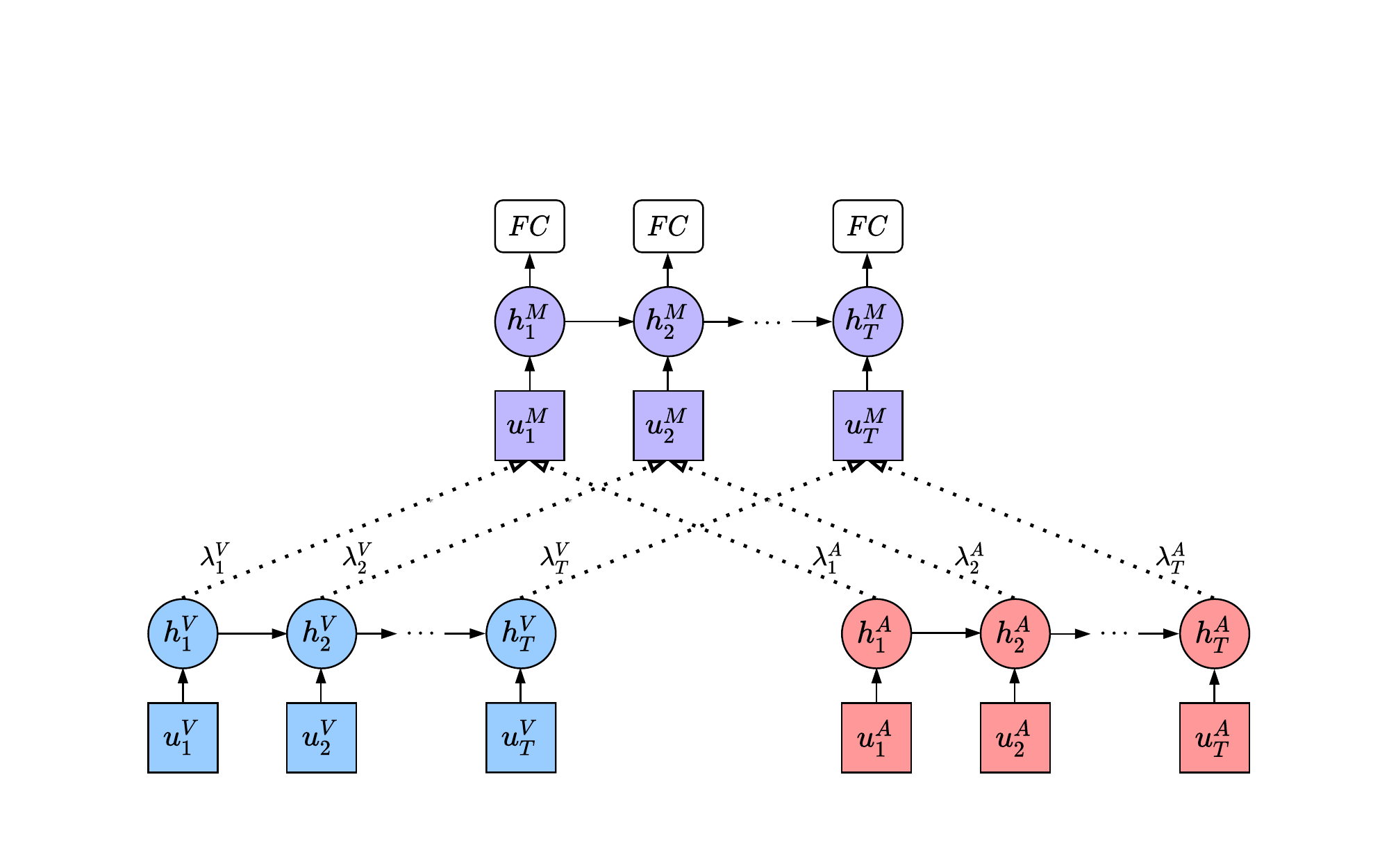}
   \caption{Multimodal fusion scheme where video and audio embedded representations are merged. FC represents a 2-dim Fully Connected layer with a softmax function.}~\\[-5ex]
\label{fig:fusion}
\end{figure}

This section describes a multimodal fusion that focuses on disentangling ambiguous scenarios in ASD problem with self-attention and uncertainty estimation. The key strategy here is to consider dynamically varying weights for each modality at any given time by computing an estimate of their relative uncertainty. Fig.\ref{fig:fusion} illustrates the concept of our fusion mechanism, inspired by the method presented in \cite{melissa}, that 
fuses the information from both modalities in a hierarchical fashion. First, we feed each video $(u_t^V)_{t=1}^T$ and audio $(u_t^A)_{t=1}^T$ embedded sequences to separated 128-dim BiGRU layers with hidden state $h_{t}^{\{{V,A}\}}$. 
Then, we compute the multimodal embedding $u_t^M$ using weighted concatenation as given by Eq.\ref{concatenation}:
\begin{equation}
u_t^M = \lambda_t^V h_t^V \oplus \lambda_t^A h_t^A
\label{concatenation}
\end{equation}
where the weights $\lambda_t^V$ and $\lambda_t^A $ characterise, at each time $t$, the uncertainty of the video and audio modalities. Finally, we use post-fusion 100-dim BiGRU layers with hidden state $h_{t}^M$, a \mbox{2-dimensional} Fully Connected (FC) layer, and a softmax.

In \cite{fusion_before_melissa, melissa}, the $\lambda$ weights are learnt using perceptrons that fully-connect, at any given time, the hidden representation of each modality.
This paper presents a novel approach to compute the weights of the modalities using: (1) an estimation of the uncertainty of both video and audio embedded representations, and (2) self-attention to measure the importance of video and audio modalities in their local temporal context.\\

\noindent{\bf High-Level Attributes:} To characterize the uncertainty, we first rely on a high-level assessment of the quality of the video and audio data. We associate each multimodal training sample $M_t^n$ with two attributes $\eta_t$ and $\mu_t^n$ that characterize the number of detected faces in frame $X_t$ and the resolution of the targeted face $x_t^n$, respectively. Indeed, higher the number of potential speakers lower the chance for the audio to be discriminative. Similarly, low-resolution face coincides with prediction uncertainty as it is more difficult to interpret facial cues and lips movements on a smaller face thumbnail. We denote $\eta_t = card(X_t)$ and $\mu_t^n$ the number of pixels in the face thumbnail $x_t^n$ before any resizing. To scale the two quantities $\eta$ and $\mu$, we first use data binning then target encoding \cite{target_encoding} to replace each value with a blend of posterior and prior probabilities of target over the entire training data. Five bins are used to pre-process the number of detected faces (four bins for $\eta\leq4$ plus a bin handling $\eta\geq5$). Face-size values are discretized within eight bins with first and last bins being left-open and right-open intervals, respectively.\\

\noindent{\bf Auxiliary Uncertainty:} We leverage auxiliary classifiers predictions to estimate the uncertainty in video and audio embeddings. Intuitively, the data leading to a highly polarized auxiliary decision can be considered as reliable. The output of a softmax layer cannot be used reliably as true probabilities \cite{uncertainty_thesis, calibration, Intriguing_uncertainty}. Therefore, we use a simple yet effective workaround called \textit{temperature scaling} \cite{calibration} to generate relevant auxiliary output scores using a calibrated softmax. We denote $\delta_t^V$ and $\delta_t^A$ the uncertainty values linked to the video and audio auxiliary predictions, respectively.  Given $\Tilde{y}_t^{\{{V,A}\}}$ the temperature-scaled probabilities of the video or audio auxiliary output, we define $\delta_t^{\{{V,A}\}}= 2(max \; \Tilde{y}_t^{\{{V,A}\}}-0.5)$.\\

\noindent{\bf Self-Attention:} We aim at using dot-product global self-attention \cite{dot_product_self_attention} mechanism to evaluate how important video and audio embedded representations are within their local temporal context. Given $H \in \mathbb{R}^{T\times128}$ as defined in Eq.\ref{H_matrix}, we compute the self-attention scores $S^{\{V,A\}} \in \mathbb{R}^{T\times T}$ using
Eq.\ref{self_attention_weights}.
\begin{equation}
H^{\{V,A\}} = [h_1^{\{V,A\}}, ..., h_T^{\{V,A\}}]
\label{H_matrix}
\end{equation}
\begin{equation}
S^{\{V,A\}} = softmax(H^{\{V,A\}}H^{{\{V,A\}}^T}+B)
\label{self_attention_weights}
\end{equation}
$B\in \mathbb{R}^{T\times T}$ is the bias matrix that limits the temporal context: the future timestamps are masked to preserve self-attention causality, and the distant past timestamps are masked too, to keep only the recent past events.
The motivation is to transform the score matrix to a one-dimensional array by assigning a scalar value to each intermediate feature. We extract the diagonal values of $S^{\{V,A\}}$ such as $a_t^{\{V,A\}} = diag(S^{\{V,A\}})_t$ is the attention scalar that characterizes $h_t^{\{V,A\}}$ within its local temporal context $[t\text{-}3,..,t]$. Here, the normalization effect of the softmax is critical, as each diagonal element will be scaled according to its relative importance within its context.\\

\noindent{\bf Attributes Combination:} Let $v$ be the combination of all the uncertainty indicators such that ${v_t = \{\eta_t, \mu_t, \delta_t^V, \delta_t^A, a_t^V, a_t^A\}}$. $\lambda_t^V$ and $\lambda_t^A$ in Eq.\ref{concatenation} are dynamically computed using Eq.\ref{lambda_calculation}:
\begin{equation}
\lambda_t^{\{V,A\}} = W^{\{V,A\}} v_t + b^{\{V,A\}}
\label{lambda_calculation}
\end{equation}
where $W^{\{V,A\}}$ and $b^{\{V,A\}}$ are trainable weights and biases, respectively, that are updated during the end-to-end training of the whole architecture. 




\begin{figure*}[th!]
\centering

\begin{subfigure}{.5\textwidth}
    \includegraphics[trim=2mm 0mm 0mm 2.5mm,clip, width=1\linewidth]{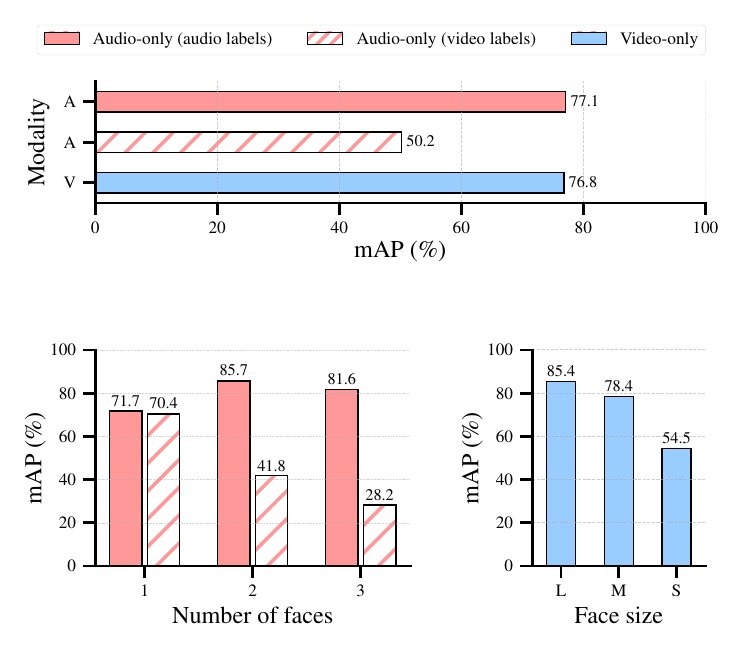}
\end{subfigure}%
\begin{subfigure}{.5\textwidth}
    \includegraphics[trim=2mm 0mm 0mm 2.5mm,clip, width=1\linewidth]{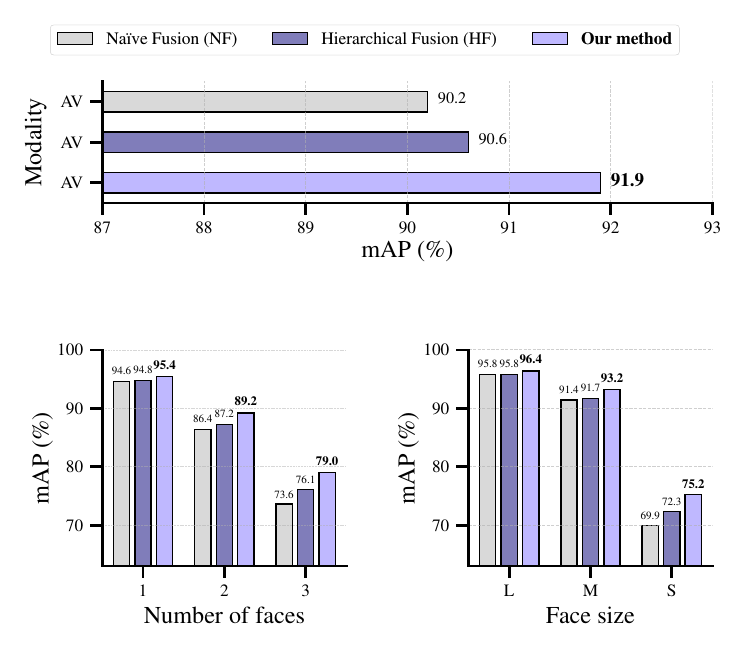}
\end{subfigure}\\[-32.75ex]
\subfloat[\label{fig5:embedding_overall} Embedding models overall performance]{\hspace{.5\linewidth}}\hspace*{0mm}
\subfloat[\label{fig5:multimodal_overall} Multimodal models overall performance]{\hspace{.5\linewidth}}\hspace*{0mm}\\[+26.5ex]
\subfloat[\label{fig5:embedding_breakdown} Embedding models performance breakdown]{\hspace{.5\linewidth}}\hspace*{0mm}
\subfloat[\label{fig5:multimodal_breakdown} Multimodal models performance breakdown]{\hspace{.5\linewidth}}\hspace*{0mm}

\caption{Performance comparison of the embedding (a,c) and the multimodal (b,d) models on the validation set. (c) and (d) present a performance analysis according to the number of detected faces (left) and the face size (right). For the number of faces, we split the validation set into three subsets by gathering one, two, and three faces frames. Altogether, these three cases cover more than 90\% of the AVA-ActiveSpeaker dataset. For the face sizes, we sort the validation dataset by ascending face-size order and split it in three equal parts denoted Small (S), Medium (M) and Large (L) respectively.}~\\[-4.5ex]
\label{fig5}
\end{figure*}

\begin{table}[t!]
\begin{center}
\caption{Performance comparison of the proposed method with the state of the art. Results are reported on both validation and ActivityNet Challenge hidden test sets.}
\label{tab:SOTA}
\begin{tabular}{lcc}
\Xhline{2\arrayrulewidth}
Method & mAP & AUC\\
\hline
\textit{Validation subset} \\
\bf{Our method} & \bf{91.9} & \bf{96.3}\\
Active Speakers Context \cite{Alcazar_2020_CVPR} & 87.1 & - \\
Huang et al. \cite{optical_flow} & - & 93.2\\
Google Baseline \cite{AVA} & 86.3 & 92.0\\
Naver Corp. (Temporal Convolutions) \cite{naver} & 85.5 & - \\
Zhang et al. \cite{Zhang2019MultiTask_3Dconv} & 84.0 & - \\
\hline
\textit{ActivityNet Challenge Leaderboard} \\
\bf{Our method} & \bf{89.5} & -\\
Naver Corp. \cite{naver} & 87.8 & -\\
Active Speakers Context \cite{Alcazar_2020_CVPR} & 86.7 & -\\
Zhang et al. \cite{Zhang2019MultiTask_3Dconv} & 83.5 & - \\
Google Baseline \cite{AVA} & 82.1 & -\\
\Xhline{2\arrayrulewidth}
\end{tabular}
\end{center}
\vspace{-8mm}
\end{table}

\section{Evaluation and Analysis}
\label{evaluation}
\subsection{Experimental Setup}
\label{evaluation:setup}

\noindent{\bf AVA-ActiveSpeaker Dataset} \cite{AVA}{\bf:} It is the most comprehensive, largest, and challenging publicly available dataset for audiovisual ASD problem. It consists of 262 movies divided into training (120), validation (33) and test (109) sets.
In total, 5,498K faces are labeled with normalized bounding boxes.
For training and validation sets, ground-truths on whether someone is speaking are also provided. The ground-truths for the test set are withheld for the annual ActivityNet Challenge \cite{activitynet_challenge}.\\

\noindent{\bf Training Strategy:} 
The ADAGRAD optimizer \cite{adagrad} is used with a learning rate of 0.015 to train the network in end-to-end fashion for 20 epochs with mini-batches of 16 sequences and without any pre-training. 
Roth et al. \cite{AVA} demonstrated that stacking few consecutive frames as input of the first 2D convolutional layer is beneficial to learn short temporal motion. Therefore, the input to the visual network (Fig.\ref{embeddings:video}) is a stack of 3 consecutive grayscale face thumbnails. The faces are extracted using the provided bounding boxes and resized to 224x224.  We feed the audio embedding network (Fig.\ref{embeddings:audio})
with 13 MFCC features extracted from the preceding 0.5s of audio with a 25ms window and a 10ms step.
Our BiGRUs are trained with 1.12s long segments (28 frames) to capture most of the different speech patterns within the AVA-Active-Speaker dataset, the average continuous speech duration being 1.11s \cite{AVA}.\\

\noindent{\bf Metrics:} For ease of comparison with previous studies, we evaluate the proposed method using the official ActivityNet Challenge evaluation script \cite{AVA} that computes the mean Average Precision (mAP) score. When available, Area Under Receiver Operating Characteristic Curve (AUC) score is also provided in Table~\ref{tab:SOTA}.
As the AVA-ActiveSpeaker test set ground-truths are kept private for the official challenge, most of our performance analysis is conducted on the validation set.

\subsection{Comparison with State Of The Art}
\label{evaluation:SOTA}

Experimental results in Table~\ref{tab:SOTA} show that the proposed method outperforms all existing approaches on both test and validation sets, in terms of mAP and AUC scores. It is worth highlighting that the proposed method outperforms one of the best approaches in \cite{Alcazar_2020_CVPR} by a significant margin of 4.8\% and 2.8\% in mAP score on the validation and test sets respectively. Furthermore, our approach surpasses Naver's method \cite{naver} by 1.7\% and ranks first in the AVA-ActiveSpeaker Leaderboard.



Table \ref{tab:numberOfParameters} compares the number of parameters of the top-ranked models on the ActivityNet Challenge Leaderboard.
Our results are very favorable since our approach has significantly fewer parameters compared to the state of the art and does not necessitates any pre-training and/or ensemble-models. 


\begin{table}[h]
\begin{center}
\caption{Comparison of the gross number of parameters.}
\label{tab:numberOfParameters}
\begin{tabular}{lcc}
\Xhline{2\arrayrulewidth}
Method & \#params & pre-training\\
\hline
\bf{Our method} & \bf{2M} & \xmark \\
Naver Corp. \cite{naver} & 13M & \cmark  \\ 
Active Speakers Context \cite{Alcazar_2020_CVPR} & 22M & \cmark  \\ 
Zhang et al. \cite{Zhang2019MultiTask_3Dconv} & 22M & \cmark  \\ 
\Xhline{2\arrayrulewidth}
\end{tabular}
\end{center}
\vspace{-8mm}
\end{table}

\subsection{Performance Breakdown} \label{evaluation:fusion_eval}

Multi-objective learning, as formulated in Section~\ref{problem_formulation}, aims at (1) learning accurate representations of each modality, and (2) allowing unbiased estimation of the uncertainty of video and audio intermediate features. We therefore evaluate the discriminative power of video and audio embedded representations. Figures \ref{fig5:embedding_overall} and \ref{fig5:embedding_breakdown} detail the performance of the embedding networks detailed in Fig.\ref{fig:embeddings} when used alone. The presented results are with 128-dim BiGRU added on top of each embedding network.
We compare the performance of the audio embedding network optimized towards either video or audio ground-truth labels as discussed in Section \ref{problem_formulation}.
As expected, the performance of the audio embedding network trained with video ground-truths strongly degrades on the number of speakers. It suffers in the ambiguous scenario presented in Fig.\ref{fig:mainPage} where multiple persons share the same audio track.
On the contrary, the performance of the audio network trained with audio labels is almost constant while increasing the number of speakers. We also observed a major mAP score increase of 26.9\% compared with using video labels. Thus,
the multi-objective approach using an independent audio-based labels allows the reliability on the audio network in difficult/ambiguous scenarios.

Figures \ref{fig5:multimodal_overall} and \ref{fig5:multimodal_breakdown} compare the performance
of our fusion method with the Naïve and Hierarchical \cite{melissa} Fusion schemes in different scenarios. Our multi-objective model is first evaluated using a naïve concatenation fusion (NF) to combine video and audio modalities. Here it is crucial to note that the method improves Active Speakers Context \cite{Alcazar_2020_CVPR} mAP by 3.1\% on validation subset (Table \ref{tab:SOTA}). This result highlights the effectiveness of our end-to-end multi-objective learning. Hierarchical Fusion (HF) refers the fusion scheme presented in \cite{melissa}. Note that our adaptation implies a slight modification of the initial method to match our "many-to-many" architecture.
As shown in Fig.\ref{fig5:multimodal_overall}, the proposed fusion scheme clearly has an advantage over NF and HF.
Fig.\ref{fig5:multimodal_breakdown} presents additional comparative analysis results by varying the number of faces detected (left) and the size of the face thumbnails (right). The proposed method clearly outperforms both NF and HF schemes, especially in challenging scenarios.

\section{Conclusion}

In this paper, we proposed a self-attention and uncertainty-based fusion mechanism that learns a comprehensive understanding of every situation towards ASD disambiguation. Our approach catered multi-objective optimization to encourage the learning of unbiased multimodal features. 
Experimental results on the challenging AVA-ActiveSpeaker dataset demonstrate
that the proposed method achieves superior performance than existing methods.
Besides, the proposed method outperformed the state-of-the-art on both validation and test datasets and ranked first in the ActivityNet Challenge, despite having fewer parameters and without any pre-training. 

\section{Acknowledgements}

This work has been supported by the French government, through the 3IA C\^ote d'Azur Investment in the Future project managed by the National Research Agency (ANR) with the reference numbers ANR-19-P3IA-0002.



\bibliographystyle{IEEEtran}

\bibliography{mybib}


\end{document}